\documentclass{WileyMSP-template}

\usepackage{mathptmx}
\usepackage{etoolbox}
\usepackage{xcolor}
\usepackage{physics}
\usepackage{bm}
\usepackage{amssymb}
\usepackage{amsmath}
\usepackage{dcolumn}
\usepackage{graphicx}

\usepackage[T1]{fontenc}
\usepackage[utf8]{inputenc}

\begin{document}

\pagestyle{fancy}
\rhead{\includegraphics[width=2.5cm]{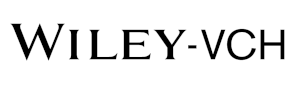}}

\title{Local droplet etching–assisted quantum dot epitaxy for telecom C-band quantum light emitters}

\maketitle


\author{Karolina E. Po{{\l}}czy{{\' n}}ska*}
\author{Pawe{{\l}} Wyborski}
\author{Micha{{\l}} Gawe{{\l}}czyk} 
\author{Shima Kadkhodazadeh} 
\author{Battulga Munkhbat} 
\author{Stefano Sanguinetti} 
\author{Elizaveta Semenova} 

\begin{affiliations}
Karolina E. Po{{\l}}czy{{\' n}}ska, Pawe{{\l}} Wyborski, Battulga Munkhbat, Elizaveta Semenova\\DTU Electro, Technical University of Denmark,
Address: \O{}rsteds Plads 343, 2800 Kongens Lyngby, Denmark\\
Email Address: karpo@dtu.dk

Shima Kadkhodazadeh\\ DTU Nanolab, Technical University of Denmark,
Address: Fysikvej 307, 2800 Kongens Lyngby, Denmark

Karolina E. Po{{\l}}czy{{\' n}}ska, Shima Kadkhodazadeh, Elizaveta Semenova\\NanoPhoton-Center for Nanophotonics, Technical University of Denmark,
Address: \O{}rsteds Plads 343, 2800 Kongens Lyngby, Denmark

Micha{{\l}} Gawe{{\l}}czyk\\Institute of Theoretical Physics, Wroc{{\l}}aw University of Science and Technology
Address: Wybrze{{\. z}}e Wyspia{{\' n}}skiego 27, 50-370 Wroc{{\l}}aw, Poland

Stefano Sanguinetti\\Department of Materials Science, University of Milano-Bicocca
Address: Via Cozzi 55, 20125 Milano, Italy

\end{affiliations}


\keywords{local droplet etching epitaxy, quantum dots, telecom C-band}

\begin{abstract}
\justify

Significant progress in quantum light sources for quantum communication applications requires reproducible and symmetric quantum emitters acting as single-photon sources capable of generating entangled photons on demand at specific telecom wavelengths. Here, we propose telecom-emitting epitaxial quantum dots (QDs) fabricated using the local droplet etching (LDE) approach. The resulting well-defined, low-density ($10^9$/cm$^2$) QDs based on In$_{x}$Ga$_{1-x}$As are formed in symmetric LDE nanoholes (in-plane aspect ratio of 1.14) in In$_{0.52}$Al$_{0.48}$As. Detailed transmission electron microscopy provides comprehensive insight into the structural integrity, interface quality, and compositional profiles of the QDs, which underpin their promising optical properties. Photoluminescence spectroscopy reveals narrow emission lines (0.2 meV) and high optical quality, while second-order autocorrelation measurements confirm clear single-photon emission, with $g^{(2)}(0)=0.07\pm0.02$ under above-band continuous-wave excitation and $g^{(2)}(0)=0.16 \pm 0.18$ under pulsed excitation.
Precise numerical modeling, combining multiband $\bm{k}\cdot\bm{p}$ and configuration-interaction methods, supports the optical characterization and identifies thermal excitation pathways that explain the persistence of emission up to liquid-nitrogen temperatures. These results highlight the versatility of the LDE approach for integrating new material systems and pave the way toward scalable fabrication of quantum light sources with tailored emission properties.

\end{abstract}
\justify

\section{Introduction}

Among solid-state quantum light sources operating in the telecom wavelength range, epitaxial quantum dots (QDs) are considered one of the leading candidates for enabling large-scale quantum communication and computing architectures \cite{HolewaRewiev2025}. In particular, applications such as quantum teleportation, quantum repeaters, and distributed quantum computing rely on sources of entangled photon pairs \cite{Stevenson2006_entangled,Dousse2010_entangled,Huber2018_entangled}. In semiconductor QDs, polarization-entangled photon pairs can be generated via the biexciton–exciton radiative cascade \cite{Akopian2006_cascade,Hafenbrak2007_cascade,Pathak2009_cascade}. A prerequisite for exploiting the biexciton-exciton cascade as a robust source of entanglement is suppressing the excitonic fine‑structure splitting (FSS) below the Fourier‑limited emission linewidth, enabling truly indistinguishable decay channels. This can only be attained in QDs with a highly symmetric in‑plane potential, underscoring the importance of synthesizing emitters with near‑ideal lateral symmetry.  The most common are QDs grown via the Stranski-Krastanov strain-driven mode \cite{Osipov2002,Li2010,Prieto2017,Berdnikov2024}, where a lattice mismatch defines the QD formation. Although this approach yields QDs with excellent optical performance, it has an inherent reduced in‑plane symmetry resulting from anisotropic indium surface diffusion during QD formation. This asymmetry leads to significant FSS. To overcome this limitation, alternative growth mechanisms have been explored, such as droplet epitaxy (DE) \cite{Gurioli2019,Holewa2022,Sala2024DEQDs}.

DE is a highly versatile technique that enables the precise tailoring of nanostructures, allowing the fabrication of a wide variety of morphologies in situ during a single epitaxial growth process \cite{Gurioli2019}. By carefully controlling parameters such as substrate temperature, annealing duration, ambient in the growth chamber, and the quantity of deposited material, it is possible to engineer nanostructures with a wide variety of shapes and properties \cite{Watanabe2000,Lee2009}, such as single rings \cite{ManoK2005}, concentric double rings \cite{Mano2005}, nanomounds \cite{Lee2006}, ring disc \cite{Somaschini2010}, and coupled dot ring structures \cite{Somaschini2011}. In addition, DE provides high control over the areal density of structures with record-low densities achieved~\cite{Wu2017}.

One branch of DE is the so‑called local droplet etching (LDE) \cite{Wang2007,Stemmann2008LDE,Heyn2015Ga,CovredaSilva2021,Cao2022,Sala2024}. LDE occurs during the annealing of metal droplets at a specific substrate temperature and atom group V ambient overpressure inside the growth chamber, such that the liquid metal droplet locally dissolves the underlying compound semiconductor layer. The dissolved material migrates through the droplet and reprecipitates at the triple phase boundary, where the liquid droplet, the solid epitaxial layer, and the gaseous ambient coexist. Simultaneously, a portion of the metal from the droplet migrates across the surface, combines with the group V atoms supplied in the chamber atmosphere, and crystallizes on the top of the surface \cite{Stemmann2008LDE}. This process results in the formation of an etched nanohole surrounded by a nanoring, which typically exhibits an elevated concentration of the droplet metal relative to the original etched epitaxial layer.

The LDE-generated nanoholes are immediately suitable for subsequent infilling with a material of choice, facilitating the formation of strain-independent QDs \cite{Heyn2009GaAsQD, Zhai2022,Babin2021,Chen2024}. By varying the amount and composition of the infilling material, the QD optical emission energy can be tuned \cite{Atkinson2012}, and complex heterostructures such as coupled nanostructures can be realized \cite{Somaschini2011}. Typically, nanoholes formed by LDE, and resulting QDs, exhibit a high degree of in-plane symmetry, compared to Stranski--Krastanov QDs, as the droplet-induced etching process is governed predominantly by the nearly isotropic dissolution of the semiconductor material into the liquid metal droplet \cite{Heyn2015,Huang2020,Shen2021}. In contrast, the pronounced elongation observed in Stranski--Krastanov QDs along specific crystallographic directions (e.g., the [1-10]) originates from anisotropic surface diffusion and facet-dependent adatom incorporation during strain-driven epitaxial growth \cite{Holewa2020}. These same anisotropies also influence the subsequent material redistribution around LDE-etched pits, giving rise to nanorings with non-ideal symmetry and distinct facets \cite{Sala2024}. However, through careful optimization of growth parameters, the in‑plane symmetry of both the nanoholes and the surrounding nanorings can be further improved. The fabrication of highly symmetric nanoholes and their subsequent filling with QD material is a promising strategy for suppressing  FSS of the excitonic states, facilitating the realization  of polarization-entangled photon emitters.

Here, we report the development of an epitaxial growth protocol and the detailed characterization of the morphology and optical properties of low-density In$_{y}$Ga$_{1-y}$As QDs obtained by filling nanoholes in  In$_{0.52}$Al$_{0.48}$ formed via LDE. The resulting QDs exhibit a characteristic two‑part morphology \cite{CovreDaSilva2026}, consisting of a highly symmetric region defined by the nanohole geometry and a slightly asymmetric dome‑shaped top that arises from material accumulation and anisotropic adatom diffusion during the infilling and overgrowth steps. We perform numerical modeling that combines multiband $\bm{k}\cdot\bm{p}$ and configuration-interaction methods based on Transmission Electron Microscope (TEM) data, to support the optical characterization. Experimentally confirmed FSS values are in a range of $40-70~$\textmu{}$eV$, comparable to the spectral resolution of the setup. Our QDs exhibit average decay times of 1.41~ns and degree of linear polarization (DOLP) of $5.43 \%$, activation energy of 4.6~meV, and $g^{(2)}(0)=0.07$, proving single-photon emission. 

\section{Morphology}

\begin{figure}[htbp]
  \includegraphics[width=\linewidth]{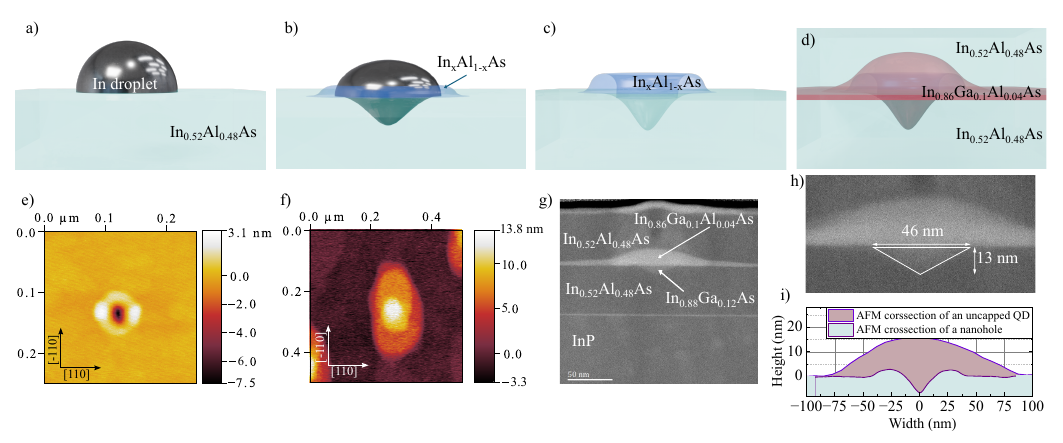}
  \caption{Schematic illustration of the QD growth based on the local droplet etching mechanism, demonstrated for our sample: a) An indium droplet is deposited on the surface of a 50~nm layer of In$_{0.52}$Al$_{0.48}$As. b) During annealing under an arsenic flux, the droplet locally melts the underlying layer, and the excess material crystallizes around the droplet, forming a nanoring. c) Once the droplet is completely consumed, the nanohole and nanoring structures are fully developed. d) The nanohole is then filled with an equivalent of 2~nm of In$_{y}$Ga$_{1-y}$As to form a QD. Finally, a 50~nm In$_{0.52}$Al$_{0.48}$As capping layer is grown. e) Nanohole etched in the In$_{0.52}$Al$_{0.48}$As layer by In droplets imaged by AFM. f) Uncapped QDs: LDE nanohole filled with 2nm of In$_{0.53}$Ga$_{0.47}$As imaged by AFM. g) HAADF STEM image of the cross-section of a QD in the sample recorded along a <110> zone-axis. The composition of each region was determined using EDS measurements (see: supplementary information, fig. S2). h) Close-up of the QD with the marked filled nanohole area measuring 46 nm in diameter and 13 nm in depth, compared with i) cross-section of an empty nanohole and an uncapped QD, measured in [110] direction with AFM, determined sizes of nanohole: 27 nm diameter, 6.6 nm depth.}
  \label{fig:morphology}
\end{figure}

The first step in LDE-based QD synthesis is the preparation of nanoholes with a controlled surface density. The LDE mechanism occurs during the annealing in the AsH$_3$ atmosphere after the deposition of metallic droplets on top of the semiconductor surface. During the development of the growth protocol, two types of droplets were tested: aluminum and indium. Despite many trials with different amounts of deposited material and annealing parameters, we have never reached the LDE regime with Al droplets. Low Al mobility indicates that high temperatures of annealing are required to observe LDE, and even after annealing at 750$^\circ$C we observed only droplets ripening. Higher temperatures would damage our sample, so we decided to proceed with another droplet material. The In droplets were deposited onto an In$_{0.52}$Al$_{0.48}$As layer lattice-matched to InP, followed by annealing in the AsH$_3$ ambient. During this annealing step, the underlying semiconductor layer is anisotropically etched. The atoms from the melted layer dissolve in the In droplet and diffuse to the surface, where they crystallize into nanorings along the droplet circumference. The whole process is illustrated in Fig.~\ref{fig:morphology}a-c.

The AFM characterization of resulting LDE nanoholes etched in the In$_{0.52}$Al$_{0.48}$As layer with a surface density of $\sim 10^{9}$ cm$^{-1}$ is presented in Fig.~S1a-c, and one exemplary case is depicted in Fig.~\ref{fig:morphology}e. Statistical data for diameters based on $2\times2$~\textmu{}m AFM images show that the mean nanohole diameter measured at the level of the plane surface is 30.5~nm in the $X=[110]$ direction and 34.8~nm in the $Y=[-110]$ direction, indicating their close to cylindrical symmetry with 1.14 aspect ratio. The high-resolution measurement of an exemplary nanohole presented in Fig.~\ref{fig:morphology}e reveals the depth of approx. 7.5~nm, enough to allow QD height tunability via the amount of deposited QD material. The cross-section (shown in Fig.~S1c) shows slight elongation of the nanohole along $[-110]$ and the asymmetry of the nanoring height, which is 30\% higher along $[-110]$ than $[110]$. Different shapes of the nanohole and the nanoring associated with different crystallographic directions are standard for the LDE mechanism observed also for different materials~\cite{Fuster2014,Shen2021,Sala2024}.

On top of the In$_{0.52}$Al$_{0.48}$As surface with etched nanoholes, we applied an equivalent of 2~nm (approx. 7~monolayers) of In$_{0.53}$Ga$_{0.47}$As. As shown in AFM studies of such a sample presented in Fig.~S1d-f and Fig.~\ref{fig:morphology}f, this amount of material overfilled nanoholes. As a result, we observe growth not only inside the nanohole, but also above and around it with an average elongation of 31\% in the $[-110]$ direction. The height of uncapped QDs is 15~nm on average. The surface density of QDs is in agreement with the surface density of nanoholes ($\sim 10^{9}$ cm$^{-1}$).

As mentioned and presented in Fig.~\ref{fig:morphology}d the nanoholes were filled with nominally 2~nm of In$_{0.53}$Ga$_{0.47}$As. One could naively expect partial, flat-terminated filling of the nanoholes with the material of designed composition. However, scanning transmission electron microscopy (STEM) characterization shown in Figs.~\ref{fig:morphology}g-h, combined with energy-dispersive X-ray spectroscopy (EDS) analysis presented in Fig.~S2, revealed a different shape and precise composition of the QDs. Nanoholes were completely filled and overgrown with In$_{0.88}$Ga$_{0.12}$As, indicating preferential In incorporation in the nanohole vicinity. The area surrounding the nanohole and corresponding to the nanoring consists of 4\% Al (In$_{0.86}$Ga$_{0.1}$Al$_{0.04}$As), indicating intermixing with the barrier and the QD material in that region. The 7 monolayer layer is uniformly distributed, and EDS analysis indicates it is composed of In$_{0.55}$Ga$_{0.28}$Al$_{0.18}$As. In this case, the unexpected Al presence was probably registered as a result of the beam broadening in the EDS data. 

Comparing STEM imaging (Fig.~\ref{fig:morphology}h) of capped QDs with AFM of empty nanoholes and uncapped QDs (Fig.~\ref{fig:morphology}i), we notice size differences. The size of the filled nanohole visible in the STEM image is almost twice that of the empty nanohole measured under AFM. Such a difference in those spatial parameters is mainly due to limitations in AFM measurements related to tip size, resulting in underestimated sizes of deep, narrow nanoholes. Another reason may be the intermixing between In$_{0.52}$Al$_{0.48}$As layer and In$_{x}$Ga$_{1-x}$As filling material that could effectively broaden the nanostructure.

Although the growth conditions employed in this study were optimized to support the formation of a uniform 2~nm layer of In$_{0.53}$Ga$_{0.47}$As on an ideally flat surface, the pre-patterned nature of our sample resulted in a significantly higher indium concentration within and above the filled nanoholes. This phenomenon can be attributed to the high mobility of indium atoms \cite{Ghazi2024}, which promotes the segregation of indium near the locally strained regions of the structure. Despite utilizing a nominally strain-free material system, it was not possible to completely eliminate the strain in the resulting nanoring structures. By design, these structures inherently incorporate a higher indium content due to indium droplet consumption during the LDE process. Consequently, indium accumulation in these regions locally modifies the lattice constant, further enhancing the strain in these areas. 

\section{Theoretical predictions}

\begin{figure}[htbp]
  \includegraphics[width=\linewidth]{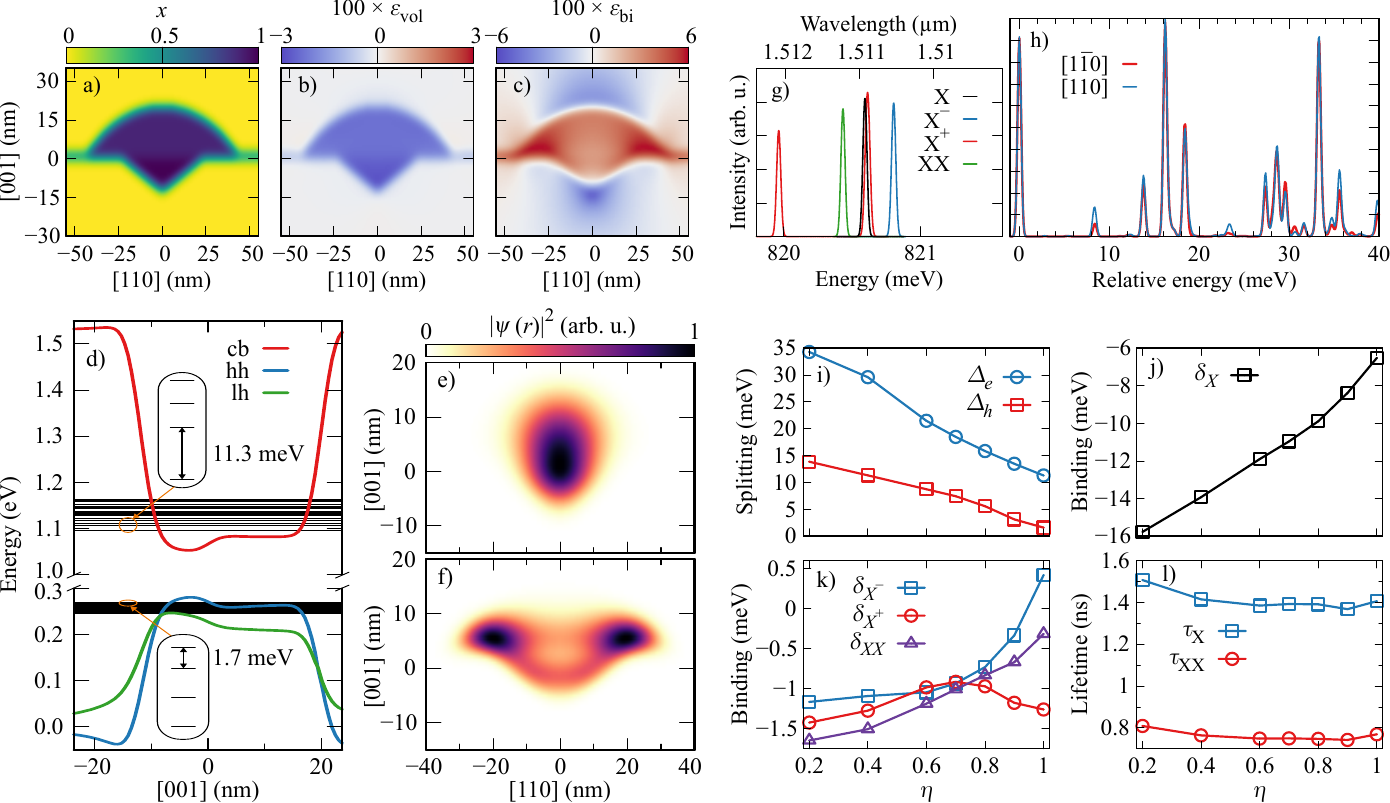}
  \caption{Theoretical predictions based on numerical calculations: Cross sections in the $(1\overline{1}0)$ plane of a) material distribution, b) volumetric strain $\varepsilon_{\mathrm{vol}}=\sum_i \varepsilon_{ii}$, c) biaxial strain $\varepsilon_{\mathrm{bi}}=2\varepsilon_{zz}-\varepsilon_{xx}-\varepsilon_{yy}$ in the simulated QD. d) Calculated single-particle energy levels (horizontal lines), shown with the band edges of the conduction and valence bands along the [001] axis (through the middle of the QD). Insets show splittings of the lowest-energy levels. e) Probability density $|\bm{\psi}|^2$ for the electron ground state. f) as in e) but for the hole. g) Calculated spectrum of the lowest-energy optical transitions arising from electron-hole recombination in excitons (black), positive (red) and negative trions (blue), and biexcitons (green). Note that two $X^{+}$ transitions fall within the range considered. h) Simulated absorption-like exciton spectrum for the two in-plane linear polarizations; The energy is given relative to the ground-state transition. i)-l) Calculated dependence on the QD dome size of i) electron (blue) and hole (red) lowest level splitting, j) exciton binding energy, k) trion and biexciton binding energies, and l) exciton and biexciton ground state radiative lifetimes.}
  \label{fig:theory}
\end{figure}

To guide the analysis of our experimental results and get more insight, we perform a detailed simulation of a QD. Uncapped QDs likely exhibit slight differences in morphology compared to capped QDs. Therefore, we use the morphology revealed in Figs.~\ref{fig:morphology}g-h and S2. Our simplified model of a QD consists of an inverted cone (46~nm base diameter $D$, 13~nm height $H$) nominally filled with In$_{0.88}$Ga$_{0.12}$As, and a 2~nm-thick layer followed by a dome-shaped cap (80~nm base diameter, 17~nm height) of In$_{0.86}$Ga$_{0.1}$Al$_{0.04}$As, all embedded in the In$_{0.52}$Al$_{0.48}$As barrier. To account for the normal diffusion of atoms at interfaces, this nominal material distribution is subject to Gaussian averaging with a spatial extent of 2~nm. In Fig.~\ref{fig:theory}a, we show the spatial distribution of the material, expressed as the value of $x$ in the (Al$_{0.48}$In$_{0.52}$As)$_{1-x}$(Al$_{0.06}$Ga$_{0.1}$In$_{0.84}$As)$_{x}$ alloy.

In Figs.~\ref{fig:theory}b-c, we show the distribution of the calculated volumetric and biaxial strains in the QD and its surroundings. The volumetric strain is low as expected for this nominally lattice-matched material system. In turn, even low biaxial strain is important for hole confinement and heavy-light hole splitting in such large QDs. We notice its inhomogeneity within the QD, with the highest values present in the external part of the dome.

The simulated energy levels for the modeled QD are shown in Fig.~\ref{fig:theory}d together with the conduction- (cb) and valence-band (heavy-hole hh, light-hole lh) edges plotted along the growth axis through the middle of the QD. The single-particle estimate for the optical transition energy is 826.7~meV (1.5~\textmu{}m). Large QD volume results in relatively dense ladders of states with the lowest level splittings of $\Delta_e=11.3$~meV for electrons and $\Delta_h=1.7$~meV for holes. Figs.~\ref{fig:theory}e-f show the probability density for the electron and hole ground states. The electron is predominantly confined in the filled nanohole, but its wavefunction also extends vertically to the dome. On the other hand, the hole is mainly confined to the outer part of the dome, with only partial overlap with the nanohole region. This specific hole confinement is due to the aforementioned biaxial strain distribution, which, via the deformation potential, creates an additional confining potential for heavy holes.

Calculation of few-carrier states allows us to simulate the spectrum of ground-state emission for the exciton, positive and negative trion, and biexciton, shown in Fig.~\ref{fig:theory}g, relative to the ground-state exciton line at 820.11~meV (1.512~\textmu{}m). We find that formation of a negative trion is weakly unfavorable compared to an exciton, as its binding energy $\delta_{X^-}=E_{X^-}-(E_X+E_e)\approx 0.4$~meV is positive. A similar measure for the biexciton is low but negative, $\delta_{XX}=E_{XX}-2E_X\approx -0.3$~meV. We predict the positive trion to be most stable with $\delta_{X^+}=E_{X^+}-(E_X+E_h)\approx -1.25$~meV, as the large dome allows for avoidance of the two holes. Noticeably, a second peak related to the excited states of $X^{+}$ is predicted to lie very close to the exciton line.

Focusing on excitonic emission, we plot in Fig.~\ref{fig:theory}h a spectrum of optical transitions simulated by calculating 2000 lowest-energy exciton states with their lifetimes and polarizations, and broadening each transition line by $\hslash/\tau$ (Fourier-limited linewidth), and convoluting with a Gaussian with an arbitrary width of $\sigma=0.2$~meV to represent the total broadening due to laser linewidth, spectral diffusion, and phonon effects \cite{Dusanowski2014}. Bright lines are observed at approximately 16~meV, 18~meV, and 28--32~meV above the ground-state transition. These transitions originate from exciton states dominated by specific electron-hole configurations $e_i h_j$, where $i$ and $j$ denote single-particle state numbers. The peak at 16~meV is due to a state dominated by the $e_2h_3$ configuration. The 18~meV peak involves two configurations, $e_1h_2$ and $e_3h_2$, while the peaks near 30~meV arise from highly mixed superpositions of various configurations that are difficult to characterize. Such a high level of configuration mixing is a fingerprint of a weaker exciton confinement regime, where electron-hole interactions significantly influence excitonic states.

Finally, we examine the dependence of the simulated QD properties on the dome size, the most variable morphological parameter. Such a study is important, as the single QDs probed with \textmu PL are typically not the majority that yield the PL maximum, but rather slight outliers. We vary the dome height and diameter uniformly as $\eta H, \eta D$ (with $\eta=1$ reproducing the nominal modeled QD and $\eta=0$ corresponding to no dome). In Fig.~\ref{fig:theory}i, we show the dependence of single-particle splittings on $\eta$. While both carriers are affected, we observe a stronger dependence for holes, with a $\approx5.5$-fold change in the splitting already for a $1.7{\times}$ change in the dome size (from $\eta=1$ to $\eta=0.6$). This is understandable in view of the hole confinement in the dome evidenced above. Tighter confinement increases the overlap of electrons and holes and, as shown in Fig.~\ref{fig:theory}j, leads to increased exciton binding energy $\delta_X=E_X-(E_e-E_h)$, where $E_X$, $E_e$, and $E_h$ are the exciton, electron, and hole ground-state energies, respectively. While the electron-hole interaction becomes stronger, its relation to electron splitting does not change much with $\Delta_e/\delta_X\approx1.6$--2.2. At the same time, a similar ratio for the hole decreases from $\Delta_h/\delta_X\approx 4$ to $\approx 1$, showing that with decreasing dome size, the confinement changes from intermediate towards a stronger confinement regime for the exciton. Reduced volume also enhances most of the bindings of higher carrier complexes with their dependence on $\eta$ shown in Fig.~\ref{fig:theory}k. Finally, we show in Fig.~\ref{fig:theory}l that in contrast to other properties, the exciton and biexciton radiative lifetimes are very weakly changed as a function of $\eta$, which underlines the resilience of the basic optical QD properties to such morphology variation. This weak change in lifetimes can be understood as arising from an interplay between electron-hole overlap, which is maximized for a small-dome QD, and the correlation-induced enhancement of emission in a large-dome QD \cite{Gawelczyk2017}. It needs to be noted here that, as a result of finite single-particle bases used in the many-particle calculations, the binding energies can be expected to be underestimated and the lifetimes overestimated.

\section{Optical properties}

\begin{figure}[htbp]
  \includegraphics[width=\linewidth]{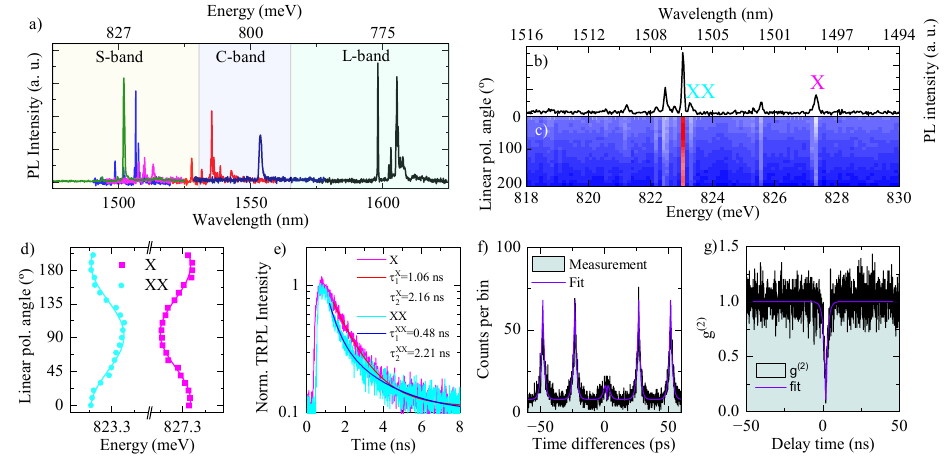}
  \caption{Optical properties of the LDE QDs. QD1: a) Microphotoluminescence spectra of single QDs observed in different spots on the sample measured at 5~K under 650~nm continuous-wave excitation. b) $\mu$PL spectrum of an individual QD with identified excitonic complexes (exciton, X, and biexciton, XX). The biexciton binding energy is 4~meV. Measurement performed at 4~K under 650~nm excitation. c) Polarization-resolved \textmu{}PL. d) Sinusoidal fit to the polarization-resolved data from panel b. FSS value obtained by the fitting: $69 \pm 9$~\textmu{}eV. e) Time-resolved photoluminescence for the X and XX lines with decay times $\tau_{1}^{X}=1.06$~ns, $\tau_{2}^{X}=2.16$~ns and $\tau_{1}^{XX}=0.48$~ns, $\tau_{2}^{XX}=2.21$~ns. QD2: f) $g^{(2)}$ measured with pulsed excitation, g) $g^{(2)}$ measured with CW excitation at 660~nm. $g^{(2)}(0)=0.07$.}
  \label{fig:optical}
\end{figure}

To verify the optical properties of QDs and their potential for single-photon emission, we performed optical measurements for the investigated sample at cryogenic temperatures using the \textmu{}PL setup (see Methods). In addition to substrate and barrier-layer emission, low-spatial-resolution measurements reveal narrow single lines in the third telecommunication window (see Fig.~S3), demonstrating the potential of this platform for realizing single-QD emitters in the telecom range. The low number of QD lines supports a low surface density of QDs, equal to that of nanoholes $\approx 10^{9}$ cm$^{-1}$, providing good access to single-QD investigations even in an unpatterned sample. More accurate high-spatial-resolution measurements reveal single-QD emission in the range between 1500~nm and 1620~nm, as shown in Fig.~\ref{fig:optical}a.

The results demonstrate a successful realization of a single-photon source in the telecom C-band using the applied epitaxial QD growth. The more representative QD analyzed in detail in Supplementary Information (Fig.~S4) shows good agreement with theoretical predictions based on the morphology revealed by STEM scans, with QDs formed by a hole and a dome. However, challenging emission line identification hinders verification of other QD properties in that case. Thus, we also focus on other QDs observed within the same structure, which are possibly less typical but show larger separations between excitonic lines.

Fig.~\ref{fig:optical}b presents the spectrum of such a QD with well-separated lines. X and XX lines are marked and show biexciton binding energy of above 4~meV. The lines are identified by using power-dependent exponent analysis, polarization-dependent FSS analysis, and decay-time-based comparison, indicating the strong confinement regime.
Moreover, the polarization-dependent PL shown in Figs.~\ref{fig:optical}c-d reveals the same FSS value of $\approx 70$~\textmu{}eV for both lines, with the expected anti-phase dependence. Fig.~S5 shows the results of power-dependent measurements showing power-law intensity scaling with $\approx 0.8$ exponent for the X line and 1.4 for the XX line. The ratio of around 1.8 is close to 2, expected in the strong confinement regime. Fig.~\ref{fig:optical}e shows decay times of $\approx 1.06$~ns ($\approx 0.48$~ns) for the X (XX) line with the ratio of about 2.2 in agreement with power-dependent analysis.

The presented results with higher XX binding and indications of confinement closer to the strong confinement regime for this QD, compared to our theoretical predictions, suggest a reduced dome size. Thus, we infer that such less typical QDs are associated with a smaller amount of active material accumulated within and above a nanohole, i.e., reduced or vanishing dome size. While such a difference in morphology significantly affects the confinement potential and binding energies, the exciton decay time stays at a similar level, in good agreement with theoretical predictions. The increased separation of the electron and hole states for a reduced QD dome size can also be correlated with higher quenching energies observed for other QDs showing a similar larger separation of the excitonic lines (see Fig.~S6).

Atypical QDs also reveal non-zero FSS, suggesting an increased anisotropy of the confinement potential. However, these higher FSS values ($<70$~\textmu{}eV) presented in Fig.~\ref{fig:optical}d and 40~\textmu{}eV shown in Fig.~S6 still represent an improvement compared with standard Stranski-Krastanov InP-substrate-based QDs with $>100$~\textmu{}eV FSS \cite{SkibaSzymanska2017}.

To further verify the potential of the investigated QDs for single-photon emission, in Figs.~\ref{fig:optical}f-g we show the autocorrelation function obtained using a CW 650~nm excitation laser, demonstrating single-photon emission with clear antibunching for an exemplary, less typical QD. In this case, a larger line separation allows for applying higher excitation power and obtaining an improved count rate, further enhanced by the high intensity of the used charged exciton line compared to other lines. 
This allowed for autocorrelation measurements under pulsed excitation at 80 MHz repetition rate with results in Fig.~\ref{fig:optical}f. Both experiments revealed as-measured $g^{(2)}(0)<0.07$, indicating the potential of QDs as single-photon sources, especially when integrated in photonic structures. The fitting of the autocorrelation data showed $g^{(2)}(0)=0.093 \pm 0.054$ value for CW and $g^{(2)}(0)=0.16 \pm 0.18$ for pulsed excitation. Moreover, the pulsed autocorrelation data are affected by the observed reexciation process, which shows additional photons after the first emission, thereby increasing the signal of $g^{(2)}(\tau)$ close to zero delay. The exact fitting of the $g^{(2)}(0)$ is hindered by the limited resolution of the obtained data.

Spectra obtained for QDs exhibiting small energy separations (presented in Fig.~S4) are consistent with theoretical predictions, indicating low binding energies of excitonic complexes, particularly for QDs incorporating the full-dome structure. The two dominant lines that exchange intensity with excitation power are attributed to the X-XX pair with a splitting of $\approx 0.41$~meV, while the third line corresponds to a charged exciton separated by $\approx 1.7$~meV, in agreement with simulations. Measured decay times range from 0.94 to 2.11~ns (average $\approx 1.41$~ns), slightly below theoretical values but within the typical range for high-quality QDs, indicating negligible non-radiative losses. A weak, slow component likely arises from delayed carrier feeding, with no clear evidence of additional degradation mechanisms. Arrhenius analysis reveals a low activation energy of $\approx 4.6$~meV (see Fig.~S4f), consistent with thermal population of excited hole states rather than carrier escape. Overall, the good agreement with theory provides insight into the observed behavior and suggests routes toward improved thermal stability, for example, by optimizing the filling step to increase carrier level spacing. Statistical analysis yields DOLP $\approx 5.43\%$, pointing to high in-plane symmetry and low valence-band mixing, in contrast to typical Stranski–Krastanov InP-based QDs \cite{SkibaSzymanska2017}.

\section{Conclusions}

We have demonstrated LDE-assisted epitaxial growth of low-density In$_{0.53}$Ga$_{0.47}$As QDs embedded in an\\In$_{0.52}$Al$_{0.48}$As matrix lattice-matched to InP, emitting in the telecom C-band with very high single photon purity. Structural and morphological characterization confirmed a two-section QD geometry composed of a symmetric cone-like structure based on a filled nanohole and a slightly asymmetric dome. Theoretical modeling showed that, with decreasing the dome size, the confinement evolves from an intermediate toward a stronger confinement regime for the exciton, while the radiative lifetimes of X and XX change only weakly, underlining the resilience of the basic optical QD properties to such morphological variations.

Optical experiments were performed for QDs with X-XX binding energies comparable to the linewidth, consistent with theoretical predictions, and for QDs with a larger X-XX binding energy of 4 meV, which is more suitable for FSS estimation. The FSS was found to be on the order of tens of $\mu eV$, indicating that further optimization of growth conditions combined with post-growth electrical or mechanical tuning will be required to achieve platform for high-quality polarization-entangled photon sources operating at telecom wavelengths based on In$_{0.53}$Ga$_{0.47}$As/In$_{0.52}$Al$_{0.48}$As system.

\section{Methods}

\subsection{Epitaxy}
The samples were grown using the TurboDisc  metal-organic vapor phase epitaxy (MOVPE) reactor using arsine (AsH$_3$), phosphine (PH$_3$), trimethylaluminum (TMAl), trimethylgallium (TMGa), and trimethylindium (TMIn) precursors with H$_2$ serving as a carrier gas. On top of the InP (100)-oriented substrate, a 500~nm InP buffer layer was grown, followed by a 50~nm lattice-matched In$_{0.52}$Al$_{0.48}$As barrier layer. The sample was then cooled from the growth temperature of 610$^{\circ}$C to 400$^{\circ}$C, and In droplets were deposited on the surface and annealed for 10~s in AsH$_3$ background. Under such conditions, the LDE takes place \cite{Fuster2014,Heyn2015,Heyn2015Ga,Heyn2016,Huang2020,Shen2021,Sala2024}. After etching, the sample was reheated to the growth temperature of 610$^{\circ}$C, and the nanoholes were filled with QD material, which was an equivalent of a 2~nm layer of In$_{0.53}$Ga$_{0.47}$As, lattice matched to the In$_{0.52}$Al$_{0.48}$As layer below and to the InP substrate, to maintain strain-free conditions. Finally, QDs were capped with 50~nm of In$_{0.52}$Al$_{0.48}$As.

\subsection{Microscopy}
Transmission Electron Microscopy (TEM) imaging was performed using a Thermo Fisher Spectra Ultra (S)TEM instrument equipped with a probe aberration corrector and operated at acceleration voltage of 300 kV. The concentration of different elements was determined from X-ray energy-dispersive spectroscopy (EDS) data collected using the Ultra EDS system. Dark-field high-angle annular dark-field Scanning TEM (HAADF STEM) images were acquired with a probe convergence semi-angle of 30~mrad and a collection semi-angle of approximately 50~mrad. The HAADF STEM and EDS data were processed and analyzed using Velox from Thermo Fisher Scientific (Version 3.14.0) and DigitalMicrograph from Gatan (Version 3.60.4435.0) software. \par
Atomic Force Microscope (AFM) imaging was performed in PeakForce Tapping mode using the Bruker AFM Dimension Icon-Pt microscope equipped with the Tap150Al-G Soft Tapping Mode AFM Probe by BudgetSensors. 

\subsection{Optical experiments}
The samples were optically characterized using a custom-built micro-photoluminescence (\textmu PL) setup. To perform low-temperature characterization, the sample was mounted inside a closed-cycle cryostat (attoDRY800xs, Attocube) operating at a base temperature of $\sim$4~K equipped with piezoelectric nanopositioners and a high-numerical-aperture telecom microscope objective (60×, NA = 0.8, LT-APO/Telecom, Attocube) integrated within the cryostat chamber. Excitation of the QDs was performed using semiconductor diode lasers (PicoQuant) at 650~nm (LDH-D-C-650), operated in continuous-wave (CW) or pulsed modes, with pulses shorter than 100~ps and adjustable repetition rates ranging from 2.5 to 80~MHz. Spectral analysis of the QDs emission was investigated using a spectrometer based on a 0.328~m focal-length monochromator (Kymera 328i, Andor, Oxford Instruments) equipped with interchangeable gratings (150 and 600~lines/mm) combined with a deep-thermo-electrically cooled (In,Ga)As linear array detector (iDus 1.7~\textmu m, Andor, Oxford Instruments), providing a maximum spectral resolution of $\sim$70~\textmu eV. Polarization-resolved \textmu PL measurements were performed using a rotating half-wave plate followed by a linear polarizer placed in the detection path. Time-correlated single-photon-counting mode was employed for time-resolved \textmu PL characterization and single-photon statistics measurements. Detection was carried out using superconducting nanowire single-photon detectors (SNSPDs; ID281, ID Quantique) connected to a high-resolution single-photon counting module (Time Tagger Ultra, Swabian Instruments). To verify single-photon emission, a Hanbury Brown and Twiss (HBT) configuration was used employing a 50:50 telecom fiber beamsplitter.

\subsection{Theoretical modeling}
To model the QD, we represent the material distribution within a computational box on a regular Cartesian mesh of points. Then, we determine the spatial distribution of the strain tensor $\varepsilon_{ij}(\bm{r}) = (\partial u_i / \partial x_j + \partial u_j / \partial x_i)/2$, with $\bm{u}$ the displacement field, by minimizing the elastic energy of the system $\int \mathrm{d}^3 r \sum_{ijkl} C_{ijkl}(\bm{r}) \varepsilon_{ij}(\bm{r}) \varepsilon_{kl}(\bm{r})$ within the framework of continuum elasticity theory, where $C_{ijkl}(\bm{r})$ are the position-dependent (due to composition variations) elastic stiffness constants of the material. Although the system is weakly strained, we calculate the shear-strain-induced piezoelectric polarization  $\bm{P}(\bm{r})$ up to second-order terms in the strain tensor elements, $P_i = \sum_{j} e_{ij}\varepsilon_{j} + \tfrac{1}{2}\sum_{jk} B_{ijk}\varepsilon_{j}\varepsilon_{k}$, where Voigt notation is used for the strain components, and $e_{ij}$ and $B_{ijk}$ denote the first- and second-order piezoelectric tensors, respectively. $\bm{P}(\bm{r})$ gives rise to an inhomogeneous electrostatic potential $\varphi(\bm{r})$ obtained by solving Poisson’s equation, $\nabla \cdot [\epsilon_0 \epsilon_r(\bm{r}) \nabla \varphi(\bm{r})] = -\nabla \cdot \bm{P}(\bm{r})$.

The composition profile, strain, and piezoelectric potential are then used to calculate electron and hole eigenstates employing the custom implementation \cite{Gawarecki2014,Mielnik2018} of the 8-band $\bm{k}\cdot\bm{p}$ method in the envelope function approximation \cite{Bahder1992}, also including spin-orbit effects. The explicit Hamiltonian used can be found in Ref.~\cite{Mielnik2018}, while material parameters used are listed in Ref.~\cite{Gawelczyk2017} and were collected from Refs.~\cite{VurgaftmanJAP2001,SaidiJAP2010,TseJAP2013,AmirtharajBOOK1994}. The diagonalization of the $\bm{k}\cdot\bm{p}$ Hamiltonian yields the eigenstates as eight-component pseudospinors $\bm{\psi}(\bm{r})=\sum_{\mu}\phi_{\mu}(\bm{r})\ket*{\mu}$ of envelope functions $\phi_{\mu}(\bm{r})$ for the considered bands (conduction, heavy, light, and split-off hole bands) labeled by $\mu$. Hole states are calculated by applying the time-reversal operator to the valence-band electron states.

Further, we construct an electron-hole configuration (product) basis using 48 electron and 48 hole states, in which we calculate the exciton eigenstates, $\ket*{\text{X}_n}=\sum_{ij}c_{ij}^{(n)}\ket*{\text{e}_i}\otimes\ket*{\text{h}_j}$, with complex coefficients $c_{ij}^{(n)}$ using the configuration-interaction approach \cite{Bryant1987}. For this, we include the identical-particle direct Coulomb and exchange, as well as phenomenological electron-hole exchange interactions. Finally, we evaluate within the dipole approximation the optical transition dipole moments $\bm{d}_n=e/(m_0 \omega_n)\,\bra*{\varnothing}\bm{p}\ket*{\text{X}_n}$ with $\ket{\varnothing}$ denoting the empty QD state, $\bm{p} = m_0\hslash^{-1}\,\nabla_{\bm{k}} H(\bm{k})$, and the resulting recombination times $\tau_{n} = 3\pi \varepsilon_0 \hslash c^3/(n_r  \omega_n^3 |\bm{d}_n|^{2})$, where $\omega_n = E_n/\hslash$ is the transition angular frequency, and $n_r$ is the refractive index of the surrounding medium. States of other carrier complexes and their optical transition dipoles are calculated similarly, e.g., biexciton states are calculated in the four-particle configuration basis as $\ket*{\text{XX}_m}=\sum_{ijkl}c_{ijkl}^{(m)}\ket*{\text{e}_i}\otimes\ket*{\text{e}_j}\otimes\ket*{\text{h}_k}\otimes\ket*{\text{h}_l}$, with dipole moments $\bm{d}_m=e\sum_n\bra*{\text{X}_n}\bm{r}\ket*{\text{XX}_m}/(m_0 \omega_{mn})$ where $\omega_{mn}=(E_m-E_n)/\hslash$.

\medskip
\textbf{Supporting Information} \par
Supporting Information is available from the Wiley Online Library or from the author.

\medskip
\textbf{Acknowledgements} \par 
The authors acknowledge financial support from the European Union's Horizon Europe Research and Innovation Programme under the project QPIC1550, grant agreement No 101135785, and the Danish National Research Foundation through NanoPhoton – Center for Nanophotonics, grant number DNRF147.
B.M. and P.W. acknowledge support from the European Research Council (ERC-StG ``TuneTMD'', grant no. 101076437; ERC-CoG ``Unity'', grant no. 865230) and Villum Fonden (project no. VIL53033), the TICRA foundation, and the Innovation Fund Denmark (QLIGHT, no. 4356-00002A).
M.\,G. acknowledges the financing of the MEEDGARD project funded within the QuantERA II Program that has received funding from the European Union's Horizon 2020 research and innovation program under Grant Agreement No. 101017733 and the National Centre for Research and Development, Poland --- project No. QUANTERAII/2/56/MEEDGARD/2024.
M.\,G. is grateful to Krzysztof Gawarecki for sharing his computational code.
Part of the calculations have been carried out using resources provided by the Wroc{\l}aw Centre for Networking and Supercomputing (M.\,G.).

\section*{Author declarations}

\subsection*{Conflict of Interest}
The authors have no conflicts of interest to disclose.
\subsection*{Author Contributions}

\textbf{Karolina E. Po{{\l}}czy{{\' n}}ska} Methodology; Investigation; Formal analysis; Visualization; Writing original draft; Writing – review \& editing; Conceptualization. 
\textbf{Pawe{{\l}} Wyborski} Investigation; Formal analysis; Writing – review \& editing.
\textbf{Micha{{\l}} Gawe{{\l}}czyk} Investigation; Visualization; Formal analysis; Writing – review \& editing.
\textbf{Shima Kadkhodazadeh} Investigation; Formal analysis.
\textbf{Battulga Munkhbat} Investigation.
\textbf{Stefano Sanguinetti} Conceptualization.
\textbf{Elizaveta Semenova} Conceptualization; Supervision; Writing - review \& editing; Resources, Funding acquisition.

\section*{Data Availability Statement}

The data that support the findings of this study are available from the corresponding author upon reasonable request.

\medskip

\bibliographystyle{MSP}
\bibliography{LDEQDs}

\end{document}


\pagestyle{fancy}
\rhead{\includegraphics[width=2.5cm]{vch-logo.png}}

\title{Local droplet etching–assisted quantum dot epitaxy for telecom C-band quantum light emitters - Supplementary Information}

\maketitle

\author{Karolina E. Po{{\l}}czy{{\' n}}ska*}
\author{Pawe{{\l}} Wyborski}
\author{Micha{{\l}} Gawe{{\l}}czyk} 
\author{Shima Kadkhodazadeh} 
\author{Battulga Munkhbat} 
\author{Stefano Sanguinetti} 
\author{Elizaveta Semenova} 

\begin{affiliations}
Karolina E. Po{{\l}}czy{{\' n}}ska, Pawe{{\l}} Wyborski, Battulga Munkhbat, Elizaveta Semenova\\DTU Electro, Technical University of Denmark,
Address: \O{}rsteds Plads 343, 2800 Kongens Lyngby, Denmark\\
Email Address: karpo@dtu.dk

Shima Kadkhodazadeh\\ DTU Nanolab, Technical University of Denmark,
Address: Fysikvej 307, 2800 Kongens Lyngby, Denmark

Karolina E. Po{{\l}}czy{{\' n}}ska, Shima Kadkhodazadeh, Elizaveta Semenova\\NanoPhoton-Center for Nanophotonics, Technical University of Denmark,
Address: \O{}rsteds Plads 343, 2800 Kongens Lyngby, Denmark

Micha{{\l}} Gawe{{\l}}czyk\\Institute of Theoretical Physics, Wroc{{\l}}aw University of Science and Technology
Address: Wybrze{{\. z}}e Wyspia{{\' n}}skiego 27, 50-370 Wroc{{\l}}aw, Poland

Stefano Sanguinetti\\Department of Materials Science, University of Milano-Bicocca
Address: Via Cozzi 55, 20125 Milano, Italy

\end{affiliations}


\keywords{local droplet etching epitaxy, quantum dots, telecom C-band}


\medskip
\justify

\begin{figure} [htbp]
\renewcommand{\thefigure}{S1}
\centering
\includegraphics[width=1\textwidth]{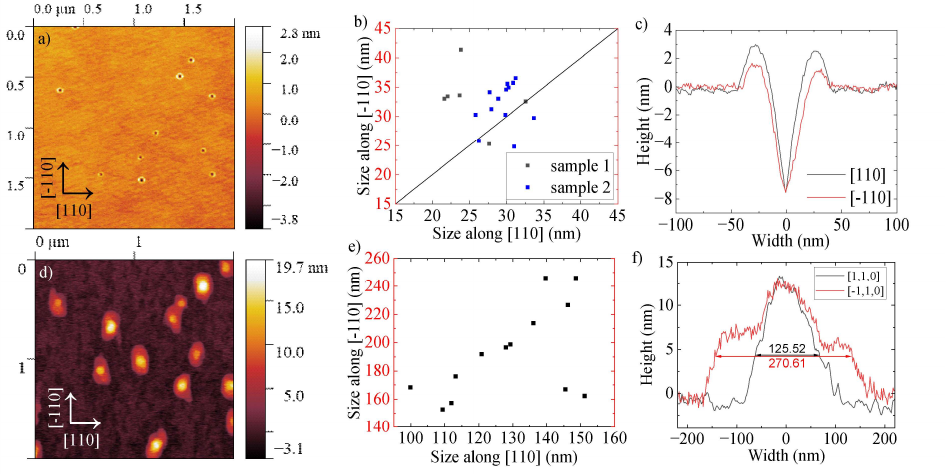}
\caption{Nanoholes etched in the In$_{0.52}$Al$_{0.48}$As layer by In droplets imaged by AFM: a) picture of the surface with etched nanoholes, b) width of nanoholes measured at the surface level along different crystallographic directions, c) cross-sections along the different crystallographic directions of the nanohole from the figure 1e from the main article. 
Uncapped QDs: LDE nanoholes filled with 2nm of In$_{0.53}$Ga$_{0.47}$As imaged by AFM: d) picture of the surface with QDs, e) width of the QDs measured at the surface level along different crystallographic directions, f) cross-sections along the different crystallographic directions of the QD from the picture figure 1f from the main article }
\label{AFM_stats}
\end{figure}

The LDE process resulted in nanoholes in the In$_{0.52}$Al$_{0.48}$As layer with a surface density of $\sim 10^{9}$ cm$^{-1}$, as presented in Fig.~\ref{AFM_stats}. AFM analysis revealed mean diameters of 30.5 nm along $[110]$ and 34.8 nm along $[-110]$, corresponding to a near-cylindrical symmetry with an aspect ratio of 1.14, and an average depth of approximately 7.5 nm. The nanoholes exhibit a slight elongation and anisotropy consistent with the LDE mechanism. Deposition of 2 nm of In$_{0.53}$Ga$_{0.47}$As leads to nanohole overfilling and the formation of quantum dots with an average height of 15 nm and lateral elongation of 30\% along $[-110]$. The resulting QD density closely matches the initial nanohole density, confirming deterministic nucleation.

\begin{figure}[htbp]
\renewcommand{\thefigure}{S2}
\centering
\includegraphics[width=0.5\textwidth]{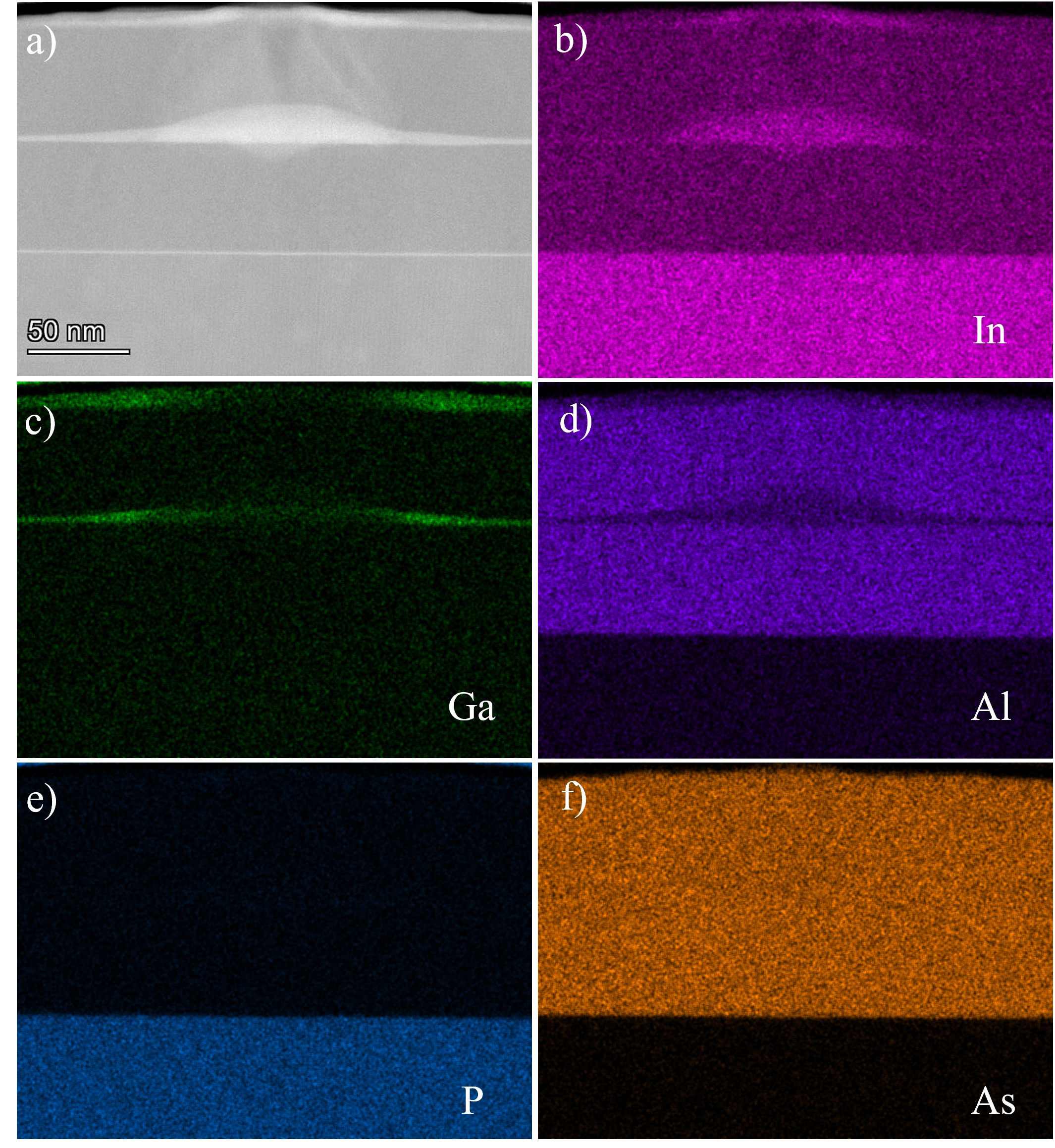}
\caption{Chemical composition of the QD and its surrounding area. a) TEM picture of the QD and its 2D elemental composition maps by EDS of b) In (magenta), c) Ga (green), d) Al (purple), e) P (blue), and f) As (orange), respectively.}
\label{TEMelements}
\end{figure}

Fig. \ref{TEMelements} presents a graphical representation of the EDS signal obtained from the sample. The elements of interest include indium, gallium, aluminum, phosphorus, and arsenic. A detailed analysis of the EDS data allowed us to determine the elemental composition of the quantum dot (In$_{0.88}$Ga$_{0.12}$As), nanoring (In$_{0.86}$Ga$_{0.1}$Al$_{0.04}$As), and the barrier (In$_{0.52}$Al$_{0.48}$As).

\begin{figure}[htbp]
\renewcommand{\thefigure}{S3}
\centering
\includegraphics[width=0.7\textwidth]{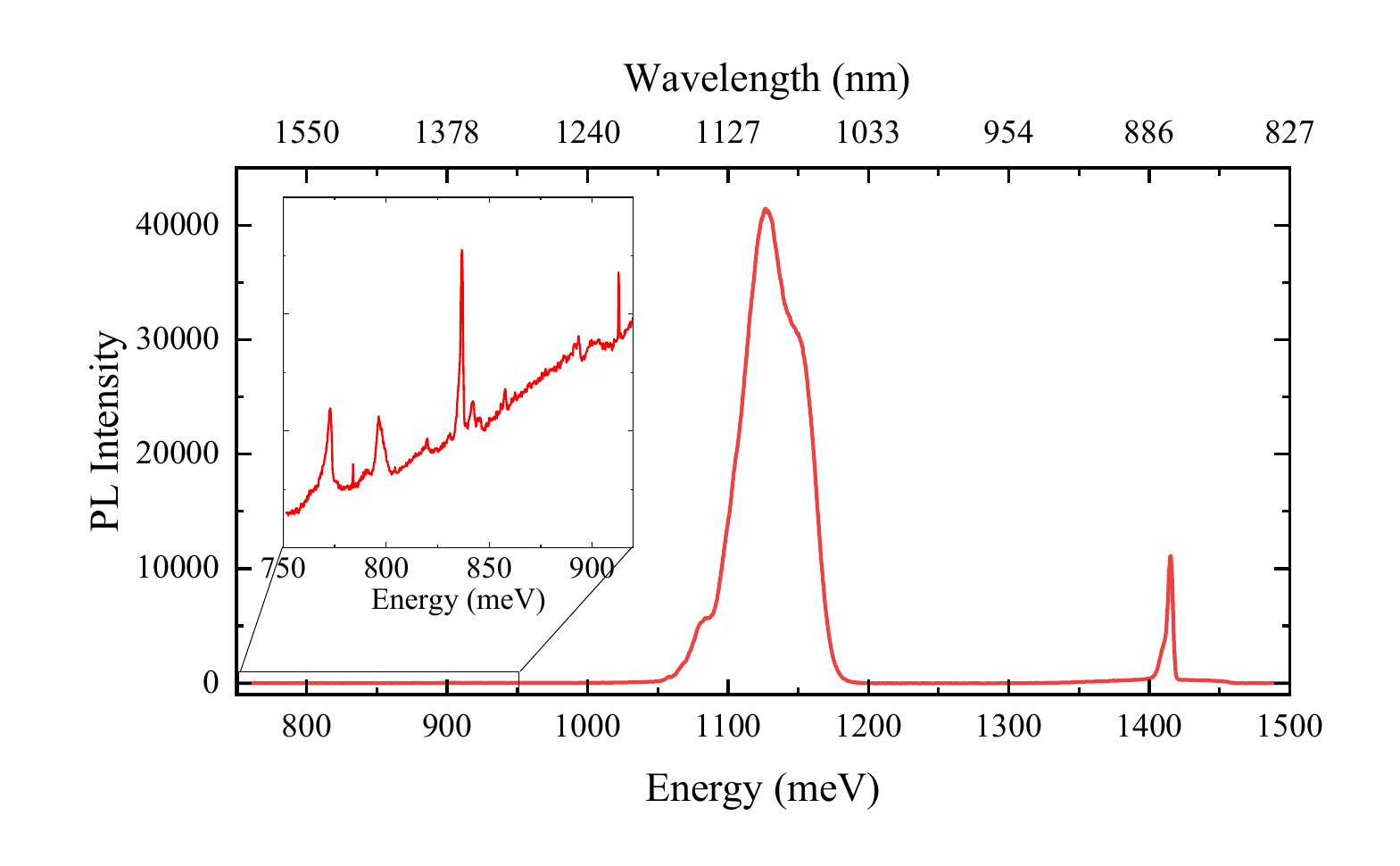}
\caption{Photoluminescence signal of the sample measured in 5 K under 650 nm CW excitation. a) macro-PL from the InP substrate at 1141 meV (1086 nm), emission related to the barrier between 1058 - 1184 meV (1047-1172nm) and single QDs emission around 752-950 meV (1305-1648nm). b) macro-PL from the QDs spectral region. c)-e) micro-PL of single QDs originating from different spots on the sample.}
\label{macroPL}
\end{figure}

The macro-PL spectrum from the Fig. \ref{macroPL} shows emission from the InP substrate at 1141 meV (1086 nm), barrier-related emission between 1058–1184 meV (1047–1172 nm), and single QD emission in the range of 752–950 meV (1305–1648 nm). Micro-PL measurements further confirm emission from individual quantum dots at different positions across the sample.

\begin{figure}[htbp]
\renewcommand{\thefigure}{S4}
\centering
\includegraphics[width=0.75\textwidth]{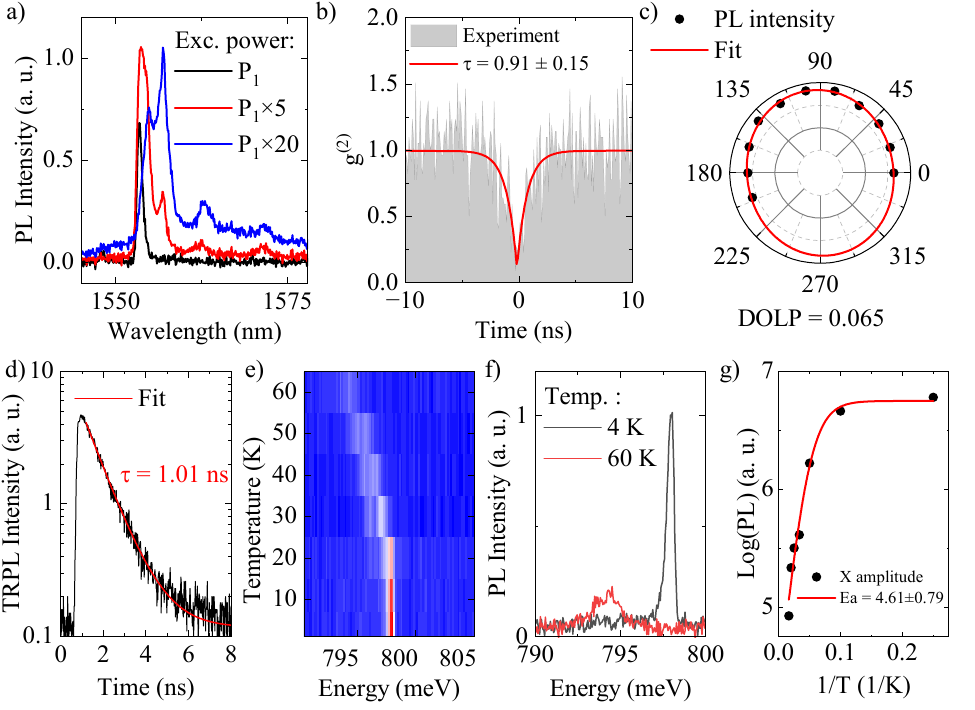}
\caption{Optical properties of the single QD. a) $\mu$PL spectrum of an individual QD measured for different power of excitation. b) g(2) measured with CW excitation of 660 nm. $g^{(2)}(0)=0.09$. c) $\mu$PL intensity as a function of the angle of polarization. DOLP = 0.065. d) Time-resolved photoluminescence, decay time from the fit $\tau_{1X}=1.01$~ns. e) $\mu$PL measured for different values of the temperature. f) comparison of the optical signal from the QD measured in 4 K and 60 K. g) Temperature dependence of the X line $\mu$PL intensity with the Arrhenius fit line. Activation energy from the fitting: $E_{a}=4.61$~meV.}
\label{theQD1}
\end{figure}

Detailed optical characterization should not only allow investigation of QD properties, but also comparison with our theoretical predictions. Fig. S3 presents the QD emission at the center of the C-band, representing the near-median of the distribution, thus expected to come from a typical QD well suited for comparison with the theoretical predictions. Low-power excitation data reveal a narrow QD line with FWHM $\approx 255$ \textmu{}eV (spectral resolution of $\approx $70~\textmu eV), suggesting still significant spectral diffusion affecting QD linewidth. Statistical analysis of 6 QDs revealed average line broadening around $200~\textmu eV$ in agreement with the presented spectrum. The two higher-power excitation spectra in Fig.~S3a show additional broadening of the main line. However, this broadening can be attributed to the presence of other lines close to the main line, as predicted theoretically for the excited state of the positive trion, which are expected to increase with the excitation power. The low separation between lines hinders particular verification of those lines' properties, specifically polarization-dependent and power-dependent properties, as well as decay time, making their identification challenging. However, the obtained spectra and their low separation are in agreement with calculation predictions, suggesting low binding energies for other excitonic complexes, especially when the QD shape includes the full size of the dome structure. Based on this agreement, we can assume the two strongest lines that exchange the intensity as the excitation power is increased to be an X-XX pair, with a splitting of $\approx 0.41$~meV, and the third visible line to be a charged exciton split from X by $\approx 1.7$~meV, which is consistent with the simulation.

Despite these challenges, we investigated main line polarization anisotropy, emission dynamics, and temperature dependence. Starting from the Hanbury Brown and Twiss configuration, we investigated single-photon emission. Fig. S4b presents the autocorrelation function obtained using CW 650~nm excitation laser, demonstrating single-photon emission with clear antibunching and a measured $g^{(2)}(\tau)~<~0.09$. The standard fitting revealed a value of $0.115~\pm~0.101$. Additionally, time-resolved PL measurements (TRPL) presented in Fig.~S4c provide insight into emission dynamics, showing a main fast decay time of around 0.96~ns combined with a rather weak slow component of around 5.7~ns. Characterization of the main decay time for other QDs revealed values ranging from 0.94~ns to 2.11~ns with an average value $\approx 1.41$~ns. These observed values are slightly lower than the theoretical prediction (see Fig.~2), consistent with the expected underestimation of the radiative rate in the calculation. These results, compared to a standard range of decay time values for InAs QDs around 1-2 ns \cite{Yang1997_InAsdec,Johansen2008_InAsdec,Hee2012}, suggest good optical quality of the investigated QDs, indicating no significant impact of non-radiative processes reducing QDs' quantum efficiency. The weak and slower decay component is likely connected to a slow channel of carrier feeding to a QD after nonresonant excitation observed for other LDE QDs \cite{Hakkarainen2024arxiv}. We cannot also exclude additional processes of carrier transfer to local charge traps. However, we do not observe other deterioration effects expected in that case.

Polarization-resolved measurements presented in Fig.~S4d show rather weak anisotropy with a degree of linear polarization (DOLP) of $6.5\pm 0.69\%$. Statistical analysis of investigated QDs reveals DOLP $\approx 5.43\%$. This highly symmetric polarization of QD emission indicates a high in-plane symmetry of the QD confinement potential (and the resulting low light-hole subband admixture to the hole ground state, especially compared with Stranski-Krastanov InP-substrate-based QDs. The fine structure splitting of the neutral excitonic transition also directly results from a reduced in-plane symmetry, and is thus indirectly connected with DOLP. Therefore, low observed DOLP values can indicate a potential for low FSS in these QDs. Direct verification of the FSS for the line in question was challenging due to its low intensity and the overlap with other lines at higher powers. However, we further analyze other QDs, showing FSS$<70$~\textmu{}eV. Together with low DOLP, this suggests the potential of the used epitaxial growth for obtaining low-FSS QDs grown on InP substrate, in contrast to asymmetric Stranski-Krastanov InP-substrate-based QDs \cite{SkibaSzymanska2017}.

To assess the suitability of QDs as quantum emitters at elevated temperatures, we also investigate their \\temperature-dependent emission. Fig. S4e presents the temperature series for the investigated main line showing a significant reduction of intensity above 10~K, however, with emission sustained up to 60~K. We observe standard redshift and line broadening due to thermal expansion and thermally enhanced carrier-phonon interaction \cite{Gammon1996,Moody2011,Ghodsi2014}. The most important and unfavorable source of emission intensity reduction is the escape of carriers from the QDs. Here, Arrhenius fitting of the main line intensity as a function of temperature shows a low characteristic energy of the emission quenching of $\approx 4.6$~meV (see Fig. S4f). Such a low energy cannot be associated with carrier escape from a hundreds-meV-deep trap. Comparing this value with theoretical predictions of the energy level separations (see Fig. 2i) suggests that the reduced intensity of the main transition is due to the promotion of the hole to excited levels within a QD. While such a process is not as critical as the carrier escape, current QDs require cryogenic temperatures to operate as a single- and indistinguishable-photon source. 
The good agreement with theoretical predictions not only explains the observed behavior but also allows us to define a clear path to more thermally stable QDs based on the same nanohole etching procedure. A refined filling step recipe preventing the accumulation of material in the dome will lead to $\approx 8$ ($\approx 3$) times larger hole (electron) level spacing.

\begin{figure}[htbp]
\renewcommand{\thefigure}{S5}
\centering
\includegraphics[width=0.5\textwidth]{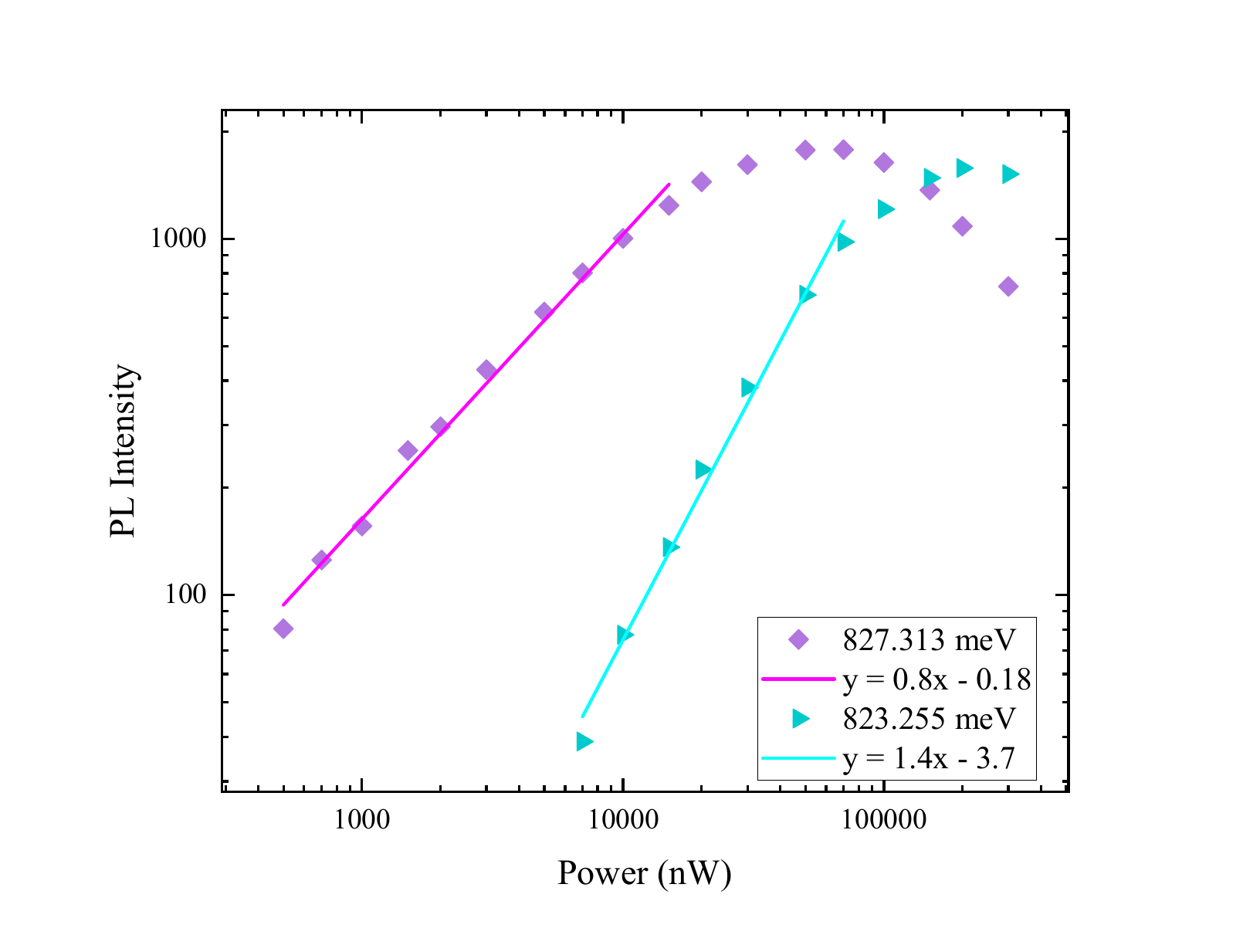}
\caption{Additional optical properties of the single QD, presented also in figs 2b-e from the main article: integrated \textmu{}PL intensity as a function of power excitation for X and XX with fit lines.}
\label{power-fig5}
\end{figure}

Fig. \ref{power-fig5} shows the results of power-dependent measurements of the QD presented also in the Fig. 3 in the main article. The Power-law intensity scales with a 0.8 exponent for the X line and 1.4 for the XX line. The ratio of around 1.8 is close to 2, expected in the strong confinement regime. 

\begin{figure}[htbp]
\renewcommand{\thefigure}{S6}
\centering
\includegraphics[width=0.75\textwidth]{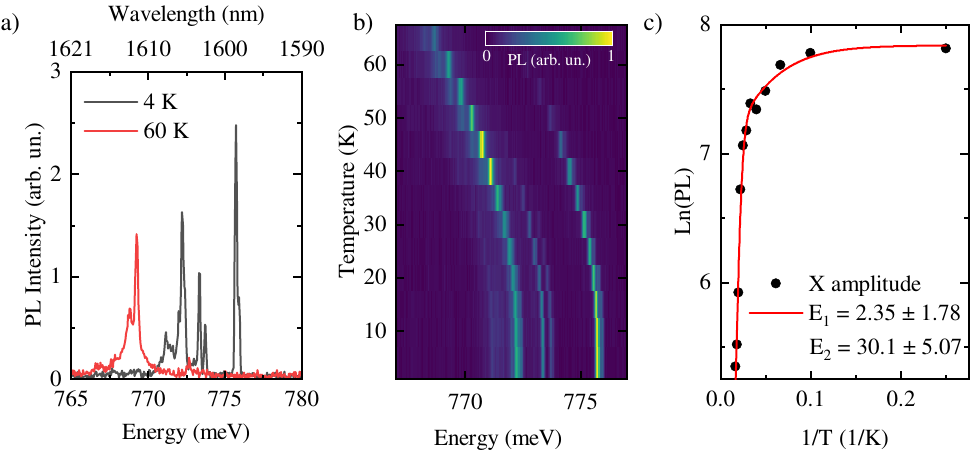}
\caption{Activation energy measurements: a) Comparison of the optical signal from the QD measured in 4 K and 60 K. b) $\mu$PL measured for different values of the temperature. c) Temperature dependence of the X line $\mu$PL intensity with the Arrhenius fit line.}
\label{Eact}
\end{figure}

An example of a QD with larger separation of the excitonic lines is shown in the Fig. \ref{Eact}. The increased separation of the electron and hole states for a reduced QD dome size can also be correlated with higher quenching energies.

\medskip

\bibliographystyle{MSP}
\bibliography{LDEQDs}